\documentclass[pra,twocolumn,nobibnotes,showpacs,amsmath,amssymb,nofootinbib]{revtex4} 

\usepackage{graphicx}   
\usepackage{dcolumn}    
\usepackage{bm}         
\newcommand{\rd}{{\rm d}} 
\newcommand{\re}{{\rm e}} 
\newcommand{\ri}{{\rm i}} 
\newcommand{\force}{f(t)}
\newcommand{\iforce}{\int_0^{t}f({t'})\,\rd{t'}}
\newcommand{\f}{a}

\newcommand{\xx}{h}
\newcommand{\ba}{\hat{b}^{\phantom\dagger}_{1}}
\newcommand{\bb}{\hat{b}^{\phantom\dagger}_{2}}
\newcommand{\bad}{\hat{b}^{\dagger}_{1}}
\newcommand{\bbd}{\hat{b}^{\dagger}_{2}}

\begin{document}


\title{Coherent control of mesoscopic tunneling
} 
 
\author{Christoph Weiss}     
\email{weiss@theorie.physik.uni-oldenburg.de}
\author{Tharanga Jinasundera}

\affiliation{Institut f\"ur Physik, Carl von Ossietzky Universit\"at, 
        D-26111 Oldenburg, Germany}

\date{\today}

\begin{abstract} 
For a weakly interacting Bose-Einstein condensate in a double well, an 
appropriate time-dependent modulation of the trapping potential
counter-acts the ``self-trapping''
effects of  the interactions, thereby allowing tunneling between the
wells.
It is demonstrated numerically
that
this modulation
can be employed for transferring the condensate from one well to the other in a controlled way.
Moreover it allows the production of mesoscopic entangled states
on short time scales. 
\end{abstract} 
 
 
\maketitle

\section{Introduction} 
Bose-Einstein condensates in double well potentials display a wide variety 
of important phenomena, ranging from interference of two condensates released 
from a double well trap~\cite{AndrewsEtAl97} and the relative phase of two 
condensates~\cite{CastinDalibard97} over the Josephson 
effect~\cite{SmerziEtAl97,RaghavanEtAl99,CataliottiEtAl01,MeierZwerger01} 
to parametric resonance~\cite{SalasnichEtAl03}. 
In this paper, we demonstrate 
that the tunneling effect between the two condensates can systematically be 
manipulated by an appropriate time-dependent variation of the double well 
potential.

For a single particle in a double well, it has been shown that the tunneling 
effect can be coherently destroyed by an external oscillating 
force~\cite{GrossmannEtAl91}. A similar effect has also been predicted for 
condensates~\cite{TsukadaEtAl99}. Here, we pursue the opposite direction:
We demonstrate numerically that under the influence of a suitably designed 
force tunneling of condensates can take place even under conditions where 
it normally is inhibited by self-trapping. 

Self-trapping has only recently been observed experimentally for condensates in a double
well potential~\cite{AlbiezEtAl05}; the ability to control tunneling in such systems will
enable the experimental realization of new phenomena.
As a particular application of our technique, we produce numerically highly entangled
mesoscopic superpositions (an entangled state cannot
be written as a product state).
While possible applications of mesoscopic entangled states include interferometry, frequency
standards and quantum information processing, the realization of mesoscopic superpositions
of condensates is still a challenge of fundamental research. So far,
superpositions of single atom states~\cite{MonroeEtAl96}, four wave 
mixing of matter waves~\cite{DengEtAl99} and even more complicated quantum 
states showing multi-particle entanglement~\cite{Mandel03} have been 
realized experimentally. There have been several suggestions to produce 
mesoscopic entanglement with condensates 
by manipulating the interaction between the particles, or by controlling the 
dynamics of the 
system~\cite{CiracEtAl98,MolmerSorensen99,SorensenEtAl01,DunninghamBurnett01,MicheliEtAl03,MahmudEtAl03}. 
The method presented here is quite simple, and robust against small 
uncertainties in both the preparation of the initial state and specific 
details of the force. It also might lend itself to a systematic 
adaption of highly successful methods developed in the field of coherent 
control of molecular dynamics~\cite{JudsonRabitz92,AssionEtAl98} to 
condensates.

Our material is organized as follows: First we introduce the $N$-particle 
model used to describe the dynamics, and 
transform the $N$-particle Schr\"odinger equation into an equivalent set of 
equations. These equations reveal the guiding principle how to design the 
required time-dependent forces (a discussion of the validity of the model can be found
in the appendix). 
We then demonstrate numerically how an 
efficient population transfer from one well to the other can be achieved, 
and outline the strategy of creating mesoscopic entanglement.

\section{\label{sec:model}Model}
For a matter-of-principle discussion, we consider the $N$-particle Hamiltonian
\begin{eqnarray}
   \hat{H} & = & -\frac{\hbar\Omega}2\left(\ba\bbd+\bad\bb\right)
   + \hbar\kappa\left(\bad\bad\ba\ba+\bbd\bbd\bb\bb\right)
\nonumber\\
\label{eq:H}
   & + & \hbar\mu\force\left(\bad\ba-\bbd\bb\right) \; ,
\end{eqnarray}
where the Bose operators~$\hat{b}^{(\dagger)}_{j}$ 
annihilate (create) one particle in the \mbox{$j$-th} well ($j=1,2$). 
In addition, $\Omega$ denotes the single-particle tunneling frequency, 
$\hbar\kappa$ is the on-site interaction energy, and $\hbar\mu f(t)$ specifies 
an externally applied potential difference between the two wells. Without this 
time-dependent modulation, this Hamiltonian is well established for describing 
the tunneling dynamics of a condensate in a double well potential within a 
two-mode approximation~\cite{ParkinsWalls98,MilburnEtAl97}. The time-dependent 
force has to be chosen such that this approximation is not violated.

Considering two harmonic wells with angular frequency~$\omega$ and oscillator 
length $\ell_{\rm osc}=\sqrt{\hbar/(2m\omega)}$ separated by a distance 
of~$2d$, conservative estimates~\cite{MilburnEtAl97,HolthausStenholm01} yield
$\Omega\simeq\omega\sqrt{2/\pi}d/\ell_{\rm osc}\exp(-0.5d^2/\ell_{\rm osc}^2)$ 
and $\kappa\simeq\hbar a/(4\sqrt{\pi}m\ell_{\rm osc}^3)$, where $m$ is the 
atomic mass and $a$ the $s$-wave scattering length. Taking $^{23}$Na atoms 
with $a=2.75\;\rm nm$, we obtain $N\kappa\simeq 1.6\Omega$ for $N=1000$ 
particles by
choosing~$\ell_{\rm osc}\simeq 31.78\rm \mu m$ and 
$d=3.2\ell_{\rm osc}$, say, with~$\omega\simeq 1.362\;$Hz, 
$\Omega \simeq0.02078\;\rm Hz$. Experimentally, time-scales shorter by two orders of magnitude can easily be  
obtained by either choosing just $N=100$ or by reducing the interaction with a Feshbach
resonance by a factor of 10. In both cases one would have: $\omega\simeq 136.2\;$Hz and
$\Omega \simeq2.078\;\rm Hz$ with $\ell_{\rm osc}\simeq3.178\rm \mu m$ and $d=3.2\ell_{\rm osc}$.


Without interaction between the particles~($\kappa=0$), and without any 
applied force~($\mu\force=0$), an initial state with all particles trapped 
in one well simply leads to Rabi-like oscillations of the entire population 
between both wells:    
$
   {\langle \bad\ba\rangle}/N = (1\pm\cos(\Omega t))/2
$.
For describing the evolution of the interacting system, we use the 
Fock basis
$
   |\nu\rangle \equiv |N-\nu,\nu\rangle 
$   
with   
$   
   \nu=0\ldots N
$, 
so that the label~$\nu$ refers to a state with $N-\nu$~particles in well~$1$,
and $\nu$~particles in well~$2$. The Hamiltonian~(\ref{eq:H}) then becomes a 
sum of two $(N+1)\times(N+1)$-matrices,
\begin{equation}
\label{eq:Hsum}
   H = H_0(t) + H_1 \; .
\end{equation}
The non-diagonal matrix~$H_1$ describes the tunneling, 
\begin{eqnarray}
   \langle \nu | H_1 | n \rangle = 
   -\left(\delta_{\nu,n-1} + \delta_{\nu-1,n}\right)
   \frac{\hbar\Omega}2 \sqrt{N-\nu}\sqrt{\nu+1} \; ,
\end{eqnarray}
whereas the diagonal matrix~$H_0$ includes both the interaction between 
the particles and the applied force,
\begin{eqnarray}
   \langle \nu | H_0(t) |n \rangle & = & 
   \delta_{\nu,n} \left[(N-\nu)(N-\nu-1) + \nu(\nu-1)\right]\hbar\kappa
\nonumber\\
   & + & \delta_{\nu,n} (N-2\nu)\hbar\mu\force \; .
\end{eqnarray}
To solve the Schr\"odinger equation
\begin{equation}
   \ri\hbar\frac{\partial}{\partial t}|\psi(t)\rangle = 
   \left(H_0(t)+H_1\right)|\psi(t)\rangle \; ,
\end{equation}
we employ the ansatz 
\begin{equation}
   \langle \nu|\psi(t)\rangle = 
   \f_\nu(t)\exp\left[-\frac{\ri}{\hbar}\int_0^t
   \langle \nu | H_0(t')|\nu\rangle\rd t'\right] \; .
\end{equation}
A few lines of simple algebra then lead to
\begin{eqnarray}
   \ri\hbar\dot{\f}_{\nu}(t) & = &
   \langle\nu | H_1 |\nu\!+\!1\rangle\xx_{\nu}(t){\f}_{\nu+1}(t)
\nonumber\\
   & + & \langle\nu|H_1|\nu\!-\!1\rangle \xx_{\nu-1}(t)^*{\f}_{\nu-1}(t) \; ,
\label{eq:equiv}
\end{eqnarray}
where we have set
$
   a_{-1}(t) \equiv a_{N+1}(t) \equiv 0 \, ,
$
and introduced the phase factors
\begin{equation}
\label{eq:phase}
   \xx_{\nu}(t) = \exp\left\{\ri\left[2(N-1-2\nu)\kappa t 
   +2\mu\iforce\right] \right \} \;.
\end{equation}
This set of differential equations is mathematically equivalent to the
$N$-particle Schr\"odinger equation governed by the Hamiltonian~(\ref{eq:H}). 
It will enable us to design force functions~$\mu f(t)$ to achieve the
desired control of tunneling. These predictions will then be tested 
numerically.

\begin{figure}
\centerline{\includegraphics[width = 0.9\linewidth, angle=0]
{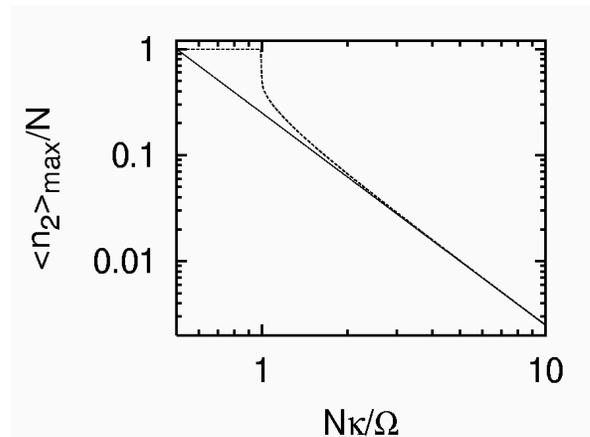}}
\caption[FIG.~1]{Maximum fraction of particles found in well~2, after 
        initially all particles had been prepared in well~1, if no
        force is applied. Solid line: Prediction from Eq.~(\ref{eq:fraction}) 
        for~$N\gg1$. Dashed line: data obtained from the solution of the 
        non-linear Schr\"odinger equation. There is a self-trapping transition 
        for $N\kappa=\Omega$~\cite{EilbeckEtAl85,KenkreCampbell86}.}
\label{fig:1}
\end{figure}

If initially all particles are stored in one well, and no force is applied 
($\mu f(t)\equiv0$), tunneling is inhibited for~$N\kappa/\Omega>1$. This 
self-trapping effect, which originally has been established on the mean-field 
level~\cite{EilbeckEtAl85,KenkreCampbell86,MilburnEtAl97,SmerziEtAl97}, has recently been
observed experimentally for condensates~\cite{AlbiezEtAl05}.
Self-trapping can also be understood within the $N$-particle approach ({\it
  cf.} Ref.~\cite{KalosakasEtAl03}): Here, we start with a 
perfectly trapped situation, in which~$\f_\nu(0)=\delta_{\nu,0}$. In the 
spirit of time-dependent perturbation theory, we then have to zeroth order 
$\f_\nu(t)=\delta_{\nu,0}$ for all times. In first order
perturbation theory, we obtain $\f_0(t)\simeq 1$,  $\f_2(t)\simeq 0$,  
$\f_3(t) \simeq 0$, \ldots, and 
\begin{equation}
   \f_1(t)\simeq\ri\frac{\Omega}2\sqrt{N}\int_0^t
   \exp\left[-\ri\,2(N-1)\kappa t'\right]\rd t' \; .
\end{equation}
Therefore, for not too large~${\Omega}/{[(N-1)\kappa]}$, the fraction of 
particles in well~1 evolves in time as
\begin{equation}
\label{eq:fraction}
   \frac{\langle\bad\ba\rangle}N \simeq 
   1 - \left(\frac{\Omega}{2(N-1)\kappa}\right)^2\sin[(N-1)\kappa t]^2 \, ,
\end{equation}
so that the transfer of population is suppressed. Surprisingly, this 
perturbative reasoning qualitatively describes self-trapping already for 
comparatively large~${\Omega}/{[(N-1)\kappa]}$; for~$N\kappa>1.5\,\Omega$ 
even the quantitative agreement is good (see Fig.~\ref{fig:1}).

\section{Transfer of Bose-Einstein condensates}
We now consider the question how to transfer the condensate from one well to 
the other even under conditions of self-trapping. If there is no interaction,
and all particles are confined in well~1 at $t=t_0$, the~$\f_\nu(t)$ which are 
essentially different from zero are centered around
$\nu_{\rm c}(t)=N\left(1-\cos[\Omega(t-t_0)]\right)/2$, 
as a consequence of the unperturbed Rabi oscillation. To reproduce this 
behavior in the interacting case, even in the self-trapping regime, the phase 
factor~$\xx_{\nu_{\rm c}}$ has to remain constant at least for small time 
differences~$\Delta t$. From Eq.~(\ref{eq:phase}), this requirement leads to 
the condition
\begin{equation}
   1 \simeq \exp\left\{\ri\big(2N\kappa\cos[\Omega (t-t_0)] 
   + 2\mu f({t})\big)\Delta t\right\} \; .
\end{equation}
To transfer a condensate initially prepared in well~1, we therefore design 
the force
\begin{equation}
\label{eq:force}
   \mu\force = \left\{\begin{array}{r@{\quad:\quad}l}
   -\mu_0 & t\le t_0\\
    \mu_1\cos[\Omega (t-t_0)] & t_0<t<t_0+\pi/\Omega \\ 
    \mu_0&t\ge t_0+\pi/\Omega \end{array} \right. \; ,
\end{equation}
with $\mu_1\approx-N\kappa$.
Fortunately, the resulting transfer is reasonably insensitive with respect
to the precise value of $\mu_1$, as the system still performs Rabi-like
oscillations for small but finite interactions. As in the experiments reported in 
Ref.~\cite{AlbiezEtAl05}, 
the role of the constant offset~$\mu_0$ is to prevent tunneling before and after the
controlled evolution of the condensate. This offset is an auxiliary device 
since it merely serves to eliminate the tunneling contact; the same effect could be achieved
by increasing the barrier.
Our numerical 
simulations show that $\mu_0$ and $\mu_1$ can be chosen such that the
experimentally uncertain value of the onset of the pulse~$t_0$ is only of 
minor importance.  We determine the optimal values for~$\mu_0$ and~$\mu_1$ 
numerically; however, slight deviations from the values given below will 
lead only to minor deteriorations of the quality of transfer.

\begin{figure}
\centerline{\includegraphics[width = 0.9\linewidth, angle=0]
{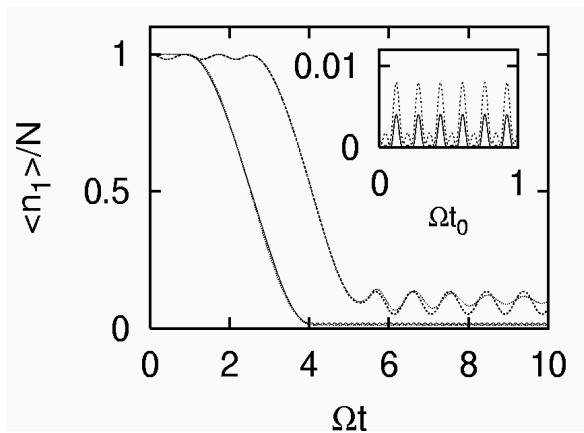}}
\caption[FIG.~2]{Population transfer as a function of time obtained for pulses 
        given by Eq.~(\ref{eq:force}). Dashed line: numerical solution of the 
        equation of motion in the mean field approximation for 
        $\mu_1 = -2.05\,\Omega$ and $\mu_0 = \mu_1$.
        Solid line: numerical solution of the mean field equation for 
        $\mu_1 = -1.8\,\Omega$ and $\mu_0 = 10\,\mu_1$.
        In both cases, the parameter $t_0$ was chosen such that we obtain 
        a worst case-scenario as far as trapping of the final state is 
        concerned ($\Omega t_0=0.92$ and $\Omega t_0=2.4$, respectively). 
        Solutions of the $N$-particle Schr\"odinger equation for~$N=1000$ 
        (dotted lines) confirm the validity of the mean field approximation.
        Inset: For the second scenario, the maximum and minimum value of 
        trapping are plotted as a function of the onset of the pulse~$t_0$ 
        (obtained from the mean field equation). The transfer to well~2
        varies between better than~$99\%$ and~$99.9\%$.}
\label{fig:trans}
\end{figure}

For the experimentally realistic parameter combination
$
   N\kappa = 1.6\,\Omega 
$
we use the Shampine-Gordon routine~\cite{ShampineGordon75} to numerically solve the
Schr\"odinger equation. The only difference between the solutions of the N-particle
Schr\"odinger equation and its mean field analog (the non-linear Schr\"odinger equation) 
is a collapse of the well-to-well oscillations on the $N$-particle level for large times 
which will be followed by a revival~\cite{MilburnEtAl97}; the N-particle dynamics displays,
on short times scales, a  
    self-trapping which is at least as good as on the mean field level.
We get good results for the transfer if we choose
$
   \mu_1 = -2.05\,\Omega
$
and
$  
   \mu_0 = \mu_1
$, 
as shown in Fig.~\ref{fig:trans}. However, neither trapping nor transfer is 
perfect then. For a very weakly coupled double well system, we can apply 
much larger constant forces. 
This will increase the effective interaction, and therefore the trapping both 
before and after the tunneling process ({\it cf.}\/ Eq.~(\ref{eq:fraction})). 
Choosing 
$
   \mu_1 = -1.8 \,\Omega
$
and
$
   \mu_0 = 10 \, \mu_1
$
leads to a transfer of more than~$99\%$ of the particles 
(see Fig.~\ref{fig:trans}). In essence, the transfer is achieved by choosing 
the force-function such that the interaction is effectively switched off for 
the dominant part of the multi-particle wave-function, according to a
neutralization of the phase factors~(\ref{eq:phase}) for the most important  
indices~$\nu$. 

\section{Generation of mesoscopic entanglement}
In what follows we exploit this possibility of effectively switching off 
the interaction to produce mesoscopic superpositions (for various measures of entanglement
see {\it e.g.\/} Ref.~\cite{Yukalov2003} and references therein). Our strategy consists
of two steps: First, an initial state has to split into two parts. Secondly, 
we separate both parts by an appropriate pulse.
The initial splitting is already achieved by starting with a state which is approximately
a Fock state~$|n\rangle$ with~$N-n$ particles in well~1 and~$n$ particles 
in well~2. For~$n=N/2$ and no applied force, the probability distribution
$|\f_{\nu}|^2$ then evolves automatically into a bimodal distribution
({\it cf. e.g.}\/ Ref.~\cite{MicheliEtAl03}), so that the state evolving 
from a Fock state already displays entanglement. A measure for the 
``quality'' of entanglement is the distance $\Delta\nu$ between the two 
maxima. Fortunately, the numerical results of entanglement generation
 are robust against small uncertainties in the preparation of the initial state.

\begin{figure}
\centerline{\includegraphics[width = 0.9\linewidth, angle=0]
{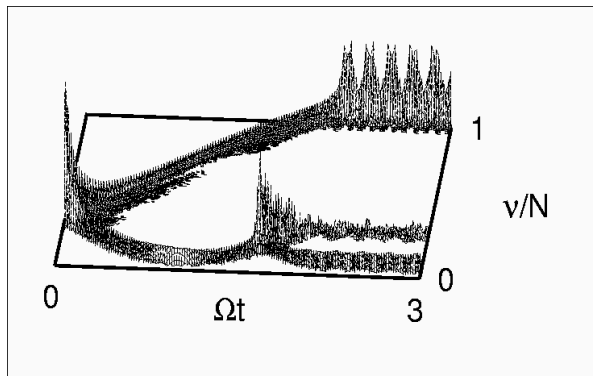}}
\caption[FIG.~3]{Mesoscopic superposition evolving from an initial 
        Fock state with~$75\%$ of the particles in well~1 and~$25\%$ in 
        well~2, under the pulse specified by Eq.~(\ref{eq:catforce}).  
        Parameters are $N\kappa=1.6\, \Omega$ and $N=1000$. 
        In this plot, probabilities $|\f_{\nu}(t)|^2$ larger than~$0.2\%$ 
        are displayed as a function of both time~$t$ and Fock state 
        index~$\nu$.}
\label{fig:cat}
\end{figure}

For a superposition of distinct Fock states, the degree of entanglement does not depend on the
phases with which the individual states contribute to the total wave-function.
To characterize such an entangled superposition more thoroughly, we perform a gedanken 
experiment by considering the outcome of all possible measurements of the 
number~$\ell$ of particles found in well~1. We then take the average of all 
measurements with~$\ell<N/2$ and call the resulting quantity
$\langle \ell\rangle_<$. Analogously, let $\langle \ell\rangle_>$ be the 
average over all measurements with~$\ell>N/2$. 
An ideal entangled state~$(|0\rangle+|N\rangle)/\sqrt{2}$ would thus be 
characterized by~$\langle \ell\rangle_<=0$ and~$\langle \ell\rangle_>=N$. 
Following the evolution of an initial Fock state numerically, we find
that $\langle \ell\rangle_> - \langle\ell\rangle_<$ oscillates in time.
For the particular initial state~$|N/2\rangle$ and $\kappa=0$, we obtain
$
   \langle \ell\rangle_> - \langle\ell\rangle_< < 0.637 \,N
$.   
Thus, even in the non-interacting case and without any applied force, 
simply letting an initial Fock state evolve will never lead to a perfect
superposition of $|N\rangle$ and $|0\rangle$. Somewhat surprisingly, for 
an interacting system in the self-trapping regime, better results can be 
obtained if we start with, say, $75\%$ of the particles in well~1 and~$25\%$ 
in well~2. Initially, roughly half of the dominant squared amplitudes 
$|\f_{\nu}|^2$ describing the many-particle wave function will move towards 
lower~$\nu$, the other half towards larger~$\nu$. To separate these two parts, 
we apply a pulse such that the interaction is effectively switched off for 
that part which shifts towards higher~$\nu$; when the other half of the 
wave-function is ``reflected'' at~$\nu=0$, the force should be such that 
trapping sets in soon. This intuitive scenario is implemented
through the force function
\begin{equation}
\label{eq:catforce}
   \mu\force = \left\{\begin{array}{r@{\quad:\quad}l}
   -N\kappa\cos[\Omega (t+t_1)] & 0\le t < \pi/\Omega-t_1 \\ 
   \mu_0 & t\ge \pi/\Omega-t_1 \end{array}\right. \; ,
\end{equation}
with
$
  \frac12(1-\cos(\Omega t_1))=\frac nN 
$.
The constant offset~$\mu_0$ is again meant to increase the trapping. 
However, now we require strong trapping both for~$\nu\approx N$ 
and~$\nu\approx 0$. Thus, a parameter $\mu_0$ of the order of~$N\kappa$ 
will not do --- either an offset with a modulus of, say, $10N\kappa$ 
can be applied, or no offset at all. Figure~\ref{fig:cat} depicts the 
evolution of the probabilities $|\f_{\nu}(t)|^2$ for finding $\nu$
particles in well~$2$ under the influence of the force~(\ref{eq:catforce}) 
with $N \kappa = 1.6\,\Omega$ and~$N = 1000$, and confirms in detail the
above qualitative reasoning: The distribution develops pronounced maxima 
at both low and high~$\nu$. For the final state, we have
$
   \langle \ell\rangle_> - \langle\ell\rangle_< > 0.73\,N\;.
$

Thus, while the final state still is no perfect entangled state, we have demonstrated
that our strategy is capable of producing states which at least come close
to the theoretical ideal. From the numerical estimates stated above for
$^{23}$Na, we infer that the production time $3/\Omega\simeq 40\;\rm ms$ 
is shorter than typical experimental coherence 
times~\cite{AlbiezEtAl05,AndrewsEtAl97,HallEtAl98}. Creating and studying mesoscopic superpositions of 
condensates might greatly improve our knowledge of the achievable degree of 
coherence in mesoscopic systems. In a laboratory experiment, the required 
initial Fock state could be prepared by involving a constant potential 
offset between the two wells, in order to prevent untimely onset of tunneling~\cite{AlbiezEtAl05}. 
Our numerical data confirm that the resulting state depends only weakly on 
the particular moment when this offset is switched off.

\section{Conclusion}

To conclude, we have shown numerically that a suitably designed time-dependent 
modulation of a double well potential can partially eliminate the effect
of the interaction between the particles of a trapped Bose-Einstein
condensate. This finding can be exploited for achieving a controlled 
tunneling transfer of the condensate from one well to the other, or for 
preparing mesoscopic entangled states. While here we have guessed the shape 
of the modulation by mere inspection of the phase factors~(\ref{eq:phase}), 
in order to establish the principal feasibility of our strategy, the result 
might be optimized if an analysis of the final state is employed, by means
of a feed-back loop, to design more efficient pulses. Thus, we suggest to 
employ the techniques which have already been developed and successfully 
applied to the coherent control of molecular dynamics by specifically designed 
laser pulses~\cite{JudsonRabitz92,AssionEtAl98} to the coherent control of 
condensates in specifically modulated double-well potentials.


\acknowledgments

We would like to thank M.~Holthaus for initiating this project and S.~Trotzky for help with
the calculations in the appendix.
Support from the DFG Priority Programme SPP~1116,
``Interactions in ultracold atomic and molecular gases'', 
is greatfully acknowledged.

\appendix

\section{validity of the two-mode approximation}
In order to demonstrate the validity of the two-mode approximation, we repeat the
mean field calculations for a non-linear Schr\"odinger equation also in a four-mode
approximation (to do the $N$-particle calculations in a four-mode approximation 
for 1000 particles exceeds the capacity
of most available computers). Throughout the appendix, we assume that the individual wells can be
described by harmonic oscillators. 
The validity of the two-mode approximation is not restricted to this situation.
However, to go beyond the two-mode approximation, assumptions about the potential have to be
made.

Starting from the non-linear Schr\"odinger equation
\begin{eqnarray}
\label{eq:nls}
\ri\hbar\frac{\partial}{\partial t}\Psi(\vec{r},t)& =& \left(-\frac{\hbar^2}{2m}\Delta + V(\vec{r})
+\vec{F}\cdot\vec{r}f(t)\right)\Psi(\vec{r},t)\nonumber\\ 
 &+& Ng|\Psi(\vec{r},t)|^2\Psi(\vec{r},t)
\end{eqnarray}
with
\begin{equation}
 g = \frac{4\pi a\hbar^2}m\;,
\end{equation}
we make the ansatz
\begin{eqnarray}
\label{eq:ansatz}
\Psi(\vec{r},t) &=& \re^{-\ri E_0t/\hbar}\left[c_1(t)u_1(\vec{r})+c_2(t)u_2(\vec{r})\right]
\\ \nonumber &+& \re^{-\ri E_1t/\hbar}\left[d_1(t)w_1(\vec{r})+d_2(t)w_2(\vec{r})\right]\;;
\end{eqnarray}
without a time-dependent potential difference ($f(t)=0$) and without interaction ($g=0$), 
$u_{1,2}$ are the single particle wave-functions for particles localized in
the ground state of
well one/two; $w_1$ and $w_2$ are the wave-functions of the first excited states.
For the weak coupling in our physical situation, the solutions of the single-particle
Schr\"odinger equation without forcing are given by 
\begin{eqnarray*}
 u_{\pm}(\vec{r}) &\simeq& \frac1{\sqrt{2}}\left[u_1(\vec{r})\pm u_2(\vec{r})\right]\\
 w_{\pm}(\vec{r}) &\simeq& \frac1{\sqrt{2}}\left[w_1(\vec{r})\pm w_2(\vec{r})\right]
 \end{eqnarray*}
with eigenenergies~$E_0^{\pm}$ for $u_{\pm}$ and $E_1^{\pm}$ for $w_{\pm}$, and
corresponding tunneling splittings:
\begin{eqnarray*}
\hbar\Omega &=& E_0^{+}-E_0^{-} \\
\hbar\Omega_1 &=& E_1^{+}-E_1^{-}\;.
\end{eqnarray*} 
After inserting Eq.~(\ref{eq:ansatz}) into Eq.~(\ref{eq:nls}) and projecting the result onto
the modes, tedious but straightforward calculations for wells which are given as
harmonic oscillators result in the non-linear Schr\"odinger equation
in the four-mode approximation: 
\begin{eqnarray}
\label{eq:four}
\ri \dot{c}_1 &=& -\frac{\Omega}2c_2  +\mu_{10}\,\re^{-\ri\omega t}f(t)d_1\\
\nonumber
 &+&\left[2 {N\kappa}|c_1|^2+4N\kappa_{10}|d_1|^2 + {\mu}f(t)\right]c_1\\
\nonumber
\ri \dot{c}_2 &=& -\frac{\Omega}2c_1 - \mu_{10}\,\re^{-\ri\omega t}f(t)d_2 \\
\nonumber
&+&\left[2N\kappa|c_2|^2 +4N\kappa_{10}|d_2|^2
 - {\mu}f(t)\right]c_2\\
\nonumber
\ri \dot{d}_1 &=& -\frac{\Omega_1}2d_2  +\mu_{10}\,\re^{-\ri\omega t}f(t)c_1\\
\nonumber
 &+&\left[2 {N\kappa_1}|d_1|^2+4N\kappa_{10}|c_1|^2 + {\mu}f(t)\right]d_1\\
\nonumber
\ri \dot{d}_2 &=& -\frac{\Omega_1}2d_1 - \mu_{10}\,\re^{-\ri\omega t}f(t)c_2 \\
\nonumber
&+&\left[2N\kappa_1|d_2|^2 +4N\kappa_{10}|c_2|^2
 - {\mu}f(t)\right]d_2\;.
\end{eqnarray}
For the interaction parameters~$\hbar\kappa=\frac g2\int |u_{1,2}|^4\,\rd^3r$,
$\hbar\kappa_1=\frac g2\int |w_{1,2}|^4\,\rd^3r$ and $\hbar\kappa_{10}=\frac g2\int
|u_{1,2}|^2|w_{1,2}|^2\,\rd^3r$ we obtain:
\begin{eqnarray*}
 \hbar \kappa_1 &=& \frac34 \hbar \kappa\\
 \hbar \kappa_{10} &=& \frac12 \hbar \kappa\;.
\end{eqnarray*}
Furthermore, we have:
\begin{equation}
\nonumber
\mu_{10} = \mu \frac{\ell_{\rm osc}}{d}
\end{equation}
(where $\frac{\ell_{\rm osc}}{d}$ is the ratio of the oscillator length and half the distance of the
two potential mimina).
The tunneling splittings are given
by~({\it cf.\/}~\cite{MilburnEtAl97,HolthausStenholm01}): 
\begin{eqnarray*}
\hbar\Omega&\simeq&\hbar\omega\sqrt{2/\pi}d/\ell_{\rm osc}\exp(-0.5d^2/\ell_{\rm osc}^2)\;, \\
\hbar\Omega_1 &=& \hbar \Omega\left(\frac{d^2}{\ell_{\rm osc}^2}-2\right)\;.
\end{eqnarray*}

\begin{figure}
\centerline{\includegraphics[width = 0.9\linewidth, angle=0]
{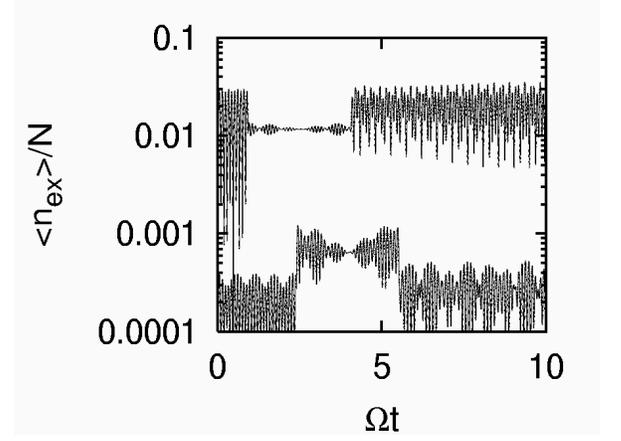}}
\caption[Ap1]{Tunneling beyond the two-mode approximation: fraction of particles which can be
found in the first two excited energy levels. The data was obtained by numerically solving
the 
non-linear
Schr\"odinger equation in four-mode approximation for the same situation as considered before
in Fig~\ref{fig:trans}.
        Top: 
        $\mu_1 = -1.8\,\Omega$ and $\mu_0 = 10\,\mu_1$. Bottom:  $\mu_1 = -2.05\,\Omega$ and
        $\mu_0 = \mu_1$.
        In both cases the parameters
        discussed in Sec.~\ref{sec:model} are used. }
\label{fig:fehler}
\end{figure}

\begin{figure}
\centerline{\includegraphics[width = 0.9\linewidth, angle=0]
{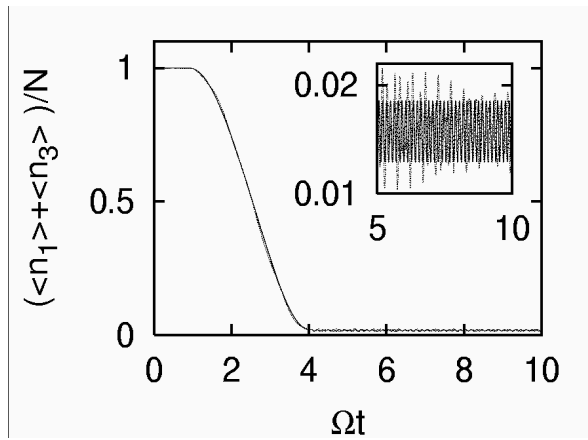}}
\caption[Ap1]{Transfer of the condensate between the wells: two-mode approximation (solid
  line) versus four-mode approximation (dotted line) for $\mu_1 = -1.8\,\Omega$ and $\mu_0 =
  10\,\mu_1$ (with the same time-dependent potential variation as in Fig.~\ref{fig:trans}) and
  the parameters discussed in Sec.~\ref{sec:model}. Although the fraction of particles
  transfered to the excited states is of the order of one percent (see top curve in
  Fig.~\ref{fig:fehler}), the  quality of the transfer between the
  wells differs only very slightly between both models (see inset).}
\label{fig:lauffour}
\end{figure}

In the derivation of Eq.~(\ref{eq:four}) we used:
\begin{eqnarray*}
  \int u_1(\vec{r})\vec{r}u_1(\vec{r}) &=&\vec{d}\\
  \int u_2(\vec{r})\vec{r}u_2(\vec{r}) &=&-\vec{d}\\
  \int w_1(\vec{r})\vec{r}w_1(\vec{r}) &=&\vec{d}\\
  \int w_2(\vec{r})\vec{r}w_2(\vec{r}) &=&-\vec{d}
\end{eqnarray*}
where $2\vec{d}$ is the vector connecting the two potential mimina.

 Within 
the two-mode approximation, coupling to the higher energy levels is neglected and 
the equations simplify to:
\begin{eqnarray}
\label{eq:two}
 \ri \dot{c}_1 = -\frac{\Omega}2c_2 + 2 {N\kappa}|c_1|^2c_1 +
 {\mu}f(t)c_1\\
\nonumber
\ri \dot{c}_2 = -\frac{\Omega}2c_1 + 2 {N\kappa}|c_2|^2c_2 -
 {\mu}f(t)c_2
\end{eqnarray}

In Fig.~\ref{fig:fehler} the fraction of particles excited to the two upper energy levels is
plotted as a function of time, for the same situation as considered before in Fig.~\ref{fig:trans}. While for the case with an initial offset of $\mu_1 =
-2.05\,\Omega$ and $\mu_0 = \mu_1$ ({\it cf.\/} Eq.~(\ref{eq:force})), the degree of
excitation lies below $0.1\%$, it is a order of magnitude larger for $\mu_1 = -1.8\,\Omega$ 
and $\mu_0 = 10\,\mu_1$. Still, even in this case the transfer between the
  wells differs only very slightly between both models (see Fig.~\ref{fig:lauffour}).

\bibliography{../langtunneling}
 
\end{document} 
%